\documentclass[conference]{IEEEtran}
\usepackage{matlab-prettifier}
\usepackage{amsmath,amssymb,amsfonts}
\usepackage{algorithmic,latexsym}
\usepackage{graphicx}
\usepackage{textcomp}
\usepackage{url,stackrel}
\usepackage{bm,float}
\usepackage{xcolor,tabu}
\usepackage{epsfig}
\usepackage{epstopdf}
\usepackage{stfloats}
\usepackage{color,empheq}
\usepackage{booktabs}
\usepackage{textgreek,tabularray}
\makeatletter

\begin{document}

\title{Design and Analysis of a Higher-Order Enhanced\\ Phase-Locked Loop via the\\ Ahmadi--Chaudhry--Zhang Newton Framework}

\author{\IEEEauthorblockN{Shafayat Abrar}
\IEEEauthorblockA{Dhanani School of Science and Engineering\\
Habib University, Karachi 75290\\
Email: shafayat.abrar@sse.habib.edu.pk}}

\maketitle

\begin{abstract}
The enhanced phase-locked loop (EPLL) is widely used in power systems to estimate the amplitude, phase, and frequency of sinusoidal voltages. Existing EPLL formulations are primarily derived from first- or second-order optimization methods, which may exhibit slow convergence, saddle-point attraction, or undesired oscillatory behavior. This paper presents a new higher-order EPLL based on the recently proposed Ahmadi--Chaudhry--Zhang (ACZ) higher-order Newton framework. Since the ACZ method was originally developed for scalar discrete-time optimization, a continuous-time higher-order Newton flow is formulated for adaptive systems. The resulting framework is then applied to the EPLL through a coordinate optimization strategy, whereby the higher-order Newton flow is applied only to the phase update, while the amplitude update retains its classical form because its cost function is exactly quadratic.

The proposed autonomous system is analyzed through phase portraits and compared with the standard, Newton, and modified EPLL formulations. Phase-portrait analysis of the autonomous model shows that the proposed flow eliminates the spurious equilibria and saddle points present in the autonomous Newton EPLL, substantially enlarges the basin of attraction associated with the desired equilibrium, and shares several desirable convergence characteristics with the modified EPLL.
\end{abstract}



\section{Introduction}

\IEEEPARstart{T}HE phase-locked loop (PLL) is essential in modern power systems, primarily for estimating line voltage parameters and tracking variations. Accurate frequency and phase information of line voltages is critical for synchronizing and controlling grid-connected power converters \cite{karimi2022modeling}. The enhanced PLL (EPLL) expands on the standard PLL by tracking sinusoidal input amplitude. First introduced in a 1982 patent \cite{mark1985residual}, the EPLL was initially described verbally and gained limited attention. Two decades later, however, Karimi-Ghartemani \textit{et al.} \cite{karimi2002nonlinear} and Wu \textit{et al.} \cite{wu2003magnitude} revisited the EPLL with a formal mathematical framework, reigniting interest in its applications.

For an input \(u(t) = \rho_n \sin \phi_n\) with nominal frequency \(\omega_n\) and voltage \(\rho_n\), the \textit{standard} EPLL (\texttt{SEPLL}) estimates \(\rho(t)\) and \(\phi(t)\) as approximations of \(\rho_n\) and \(\phi_n\), minimizing the squared error \(e(t):= u(t) - \rho(t) \sin \phi(t)\) via gradient descent. This leads to the following set of equations \cite{karimi2003periodic, KarimiBook}:
\begin{subequations}\label{EqKarimiStandardEPLL}
\begin{align}
\nonumber & \texttt{SEPLL:} \\
& \dot{\rho}(t)=\mu_1 e(t) \sin\phi(t) \\
& \dot{\omega}(t)=\mu_2 \rho(t)e(t) \cos\phi(t),\quad \omega(t_0)=\omega_n \\
& \dot{\phi}(t)=\omega(t)+\mu_3 \rho(t)e(t) \cos\phi(t),~t\geqslant t_0
\end{align}
\end{subequations}

To improve convergence, the authors in \cite{karimi2012derivation} investigated a Hessian-based second-order gradient, resulting in a \textit{Newton} EPLL implementation: 
\begin{subequations}\label{EqEPLLNewton}
\begin{align}
\nonumber & \texttt{NEPLL:} \\
& \dot{\rho}(t)=\mu_1 K(t) e(t) \sin\phi(t) \\
& \dot{\omega}(t)=\mu_2 K(t) \rho(t)^{-1}e(t) \cos \phi(t) \sin^2 \phi(t) \\ 
& \dot{\phi}(t)=\omega(t)+\mu_3 K(t) \rho(t)^{-1}e(t) \cos \phi(t) \sin^2 \phi(t)
\end{align}
\end{subequations}
where \(t\geqslant t_0\), \(\omega(t_0)=\omega_n\), and 
\begin{align*}
\frac{1}{K(t)}=\sin^2 \phi(t)\left(1+\cos^2 \phi(t)\right)-\dfrac{e(t)}{\rho(t)} \cos^2 \phi(t) \sin\phi(t)
\end{align*} 

This approach, however, introduced second-order oscillations and unintended convergence issues. The authors in \cite{karimi2012derivation} suggested an ad hoc simplification by averaging \(K(t)\), leading to the \textit{modified} EPLL formulation:
\begin{subequations}\label{EqKarimiModifiedEPLL}
\begin{align}
\nonumber & \texttt{MEPLL:} \\
& \dot{\rho}(t) =\mu_1\,e(t) \sin\phi(t) \\
& \dot{\omega}(t) =\mu_2\,\rho(t)^{-1}e(t) \cos\phi(t),\quad \omega(t_0)=\omega_n \\
& \dot{\phi}(t) =\omega(t)+\mu_3\,\rho(t)^{-1}e(t) \cos\phi(t),~t\geqslant t_0
\end{align}
\end{subequations}
The key distinction between \texttt{SEPLL} and \texttt{MEPLL} is the replacement of \(\rho(t)\) by its reciprocal in the frequency and phase updates. Although introduced heuristically, this seemingly minor modification significantly improved convergence speed and smoothness \cite{ghartemani2011problems,KarimiBook} and later enabled an equivalent linear time-invariant representation in Cartesian coordinates \cite{karimi2013linear}. A similar normalization was independently proposed by Golestan \textit{et al.} \cite{golestan2012design} for PLLs to reduce sensitivity to input-voltage amplitude variations.

In \cite{abrar2024design}, the authors established the \texttt{MEPLL} as the solution to an optimization problem through two independent derivations based on Lyapunov stability theory and natural gradient optimization. They further analyzed its convergence and stability using averaging theory and Poincaré maps, deriving explicit stability bounds for the proportional and integral gains of the loop filter. Moreover, they demonstrated that the design and tuning of the EPLL can be greatly simplified by parameterizing all three adaptation laws using a single control parameter.

Despite these advances, all existing EPLL variants are fundamentally derived from first- or second-order optimization principles. To the best of the author's knowledge, higher-order optimization methods have not yet been explored in the context of phase-locked loops. Recently, Ahmadi, Chaudhry, and Zhang proposed a higher-order Newton framework based on a convexified third-order Taylor approximation, yielding an explicit closed-form Newton update \cite{ahmadi2024higher}. Although originally developed as a discrete-time optimization algorithm, its application to continuous-time adaptive systems remains largely unexplored.

The objective of this paper is to investigate whether the Ahmadi--Chaudhry--Zhang framework can be used to design a new higher-order EPLL. Since the EPLL optimization problem naturally involves two variables, namely the signal amplitude and phase, a direct application of the scalar Ahmadi framework is not possible. We therefore adopt a coordinate optimization strategy. The higher-order Newton flow is applied only to the phase variable, whose cost function is genuinely nonlinear, while the amplitude update retains its classical second-order form because the corresponding cost is exactly quadratic and therefore contains no useful higher-order information.

The resulting algorithm is formulated as a continuous-time adaptive law for estimating the amplitude, phase, and frequency of single-phase AC signals using synthesized orthogonal components. To study its dynamical behavior, autonomous phase portraits are derived and compared with those of the standard, Newton, and modified EPLL formulations, revealing significant differences in their stationary points and basins of attraction. 

Since the Ahmadi--Chaudhry--Zhang framework is relatively recent and not yet widely known within the power electronics and control communities, this paper also includes a concise, self-contained overview of its mathematical foundations. This review establishes the notation and derivations required for the subsequent development while making the manuscript accessible to readers unfamiliar with the original work.

The contributions of this paper are summarized as follows.

\begin{enumerate}

\item A continuous-time interpretation of the Ahmadi--Chaudhry--Zhang higher-order Newton method is suggested for adaptive dynamical systems.

\item A new coordinate-wise higher-order EPLL is derived by applying the higher-order Newton flow to the phase optimization while retaining the classical update for the amplitude, whose cost function is exactly quadratic.

\item The stationary points, stability properties, and basins of attraction of the proposed autonomous dynamics are investigated through phase-portrait analysis and compared with those of the \texttt{SEPLL}, \texttt{NEPLL}, and \texttt{MEPLL}.


\end{enumerate}

\section{Phase-Portrait Analysis}

Consider a single-phase AC system where the input signal is expressed as
\begin{equation}
\label{EqUt}
u(t)=\rho_n\sin(\omega_nt+\delta_n)+u_h(t),
\end{equation}
where \(\omega_n\), \(\rho_n\), and \(\delta_n\) denote the nominal frequency, voltage magnitude, and phase angle, respectively. The term \(u_h(t)\) represents unwanted components such as DC offset, harmonics, noise, and transients. For convenience, let
\[
\phi_n=\omega_nt+\delta_n.
\]
The primary objective of the EPLL is to estimate the fundamental component of \(u(t)\), namely its amplitude, frequency, and phase angle, which are essential for accurate synchronization and control of grid-connected power electronic converters.

Since the adaptation laws of the \texttt{SEPLL}, \texttt{NEPLL}, and \texttt{MEPLL} are described by nonlinear ordinary differential equations, their convergence properties can be investigated using tools from nonlinear dynamical systems. Among these, phase-portrait analysis provides a particularly insightful graphical representation of the underlying vector field, revealing the location and nature of stationary points, their stability, basins of attraction, saddle points, and convergence trajectories. Such information is often difficult to infer directly from the governing differential equations.

Accordingly, this section investigates the stationary points and convergence characteristics of the \texttt{SEPLL}, \texttt{NEPLL}, and \texttt{MEPLL} through phase-portrait analysis. The corresponding phase portraits are generated using the MATLAB package developed in \cite{ZhangPPP}.

\subsection{Stationary Points of \texttt{SEPLL}}

The \texttt{SEPLL} in (\ref{EqKarimiStandardEPLL}) is obtained by minimizing the squared-error cost function
\(V(\boldsymbol{\theta}):\mathbb{R}^2\rightarrow\mathbb{R}_+\)
w.r.t.
\(\boldsymbol{\theta}:=[\rho,~\phi]^{\mathsf T}\).
To facilitate convergence analysis, we introduce the following autonomous cost function, which captures the essential dynamics of the original non-autonomous cost by eliminating the explicit time dependence introduced through \(u(t)\):
\begin{equation}
\label{EqVCostSEPLL2}
\begin{aligned}
V&=
\bigl(\rho\cos(\phi-\phi_n)-\rho_n\bigr)^2
+
\rho^2\sin^2(\phi-\phi_n)\\
&=
\rho^2+\rho_n^2
-
2\rho\rho_n
\cos(\phi-\phi_n),
\end{aligned}
\end{equation}
where
\(\boldsymbol{\theta}:=[\rho,~\phi]^{\mathsf T}\).
The orthogonal sine and cosine terms allow the amplitude and phase dynamics to be analyzed independently, enabling phase-portrait analysis of the autonomous system and providing insight into the convergence behavior of the original EPLL.

\begin{figure}[t!]
\centering
\includegraphics[width=0.73\linewidth]{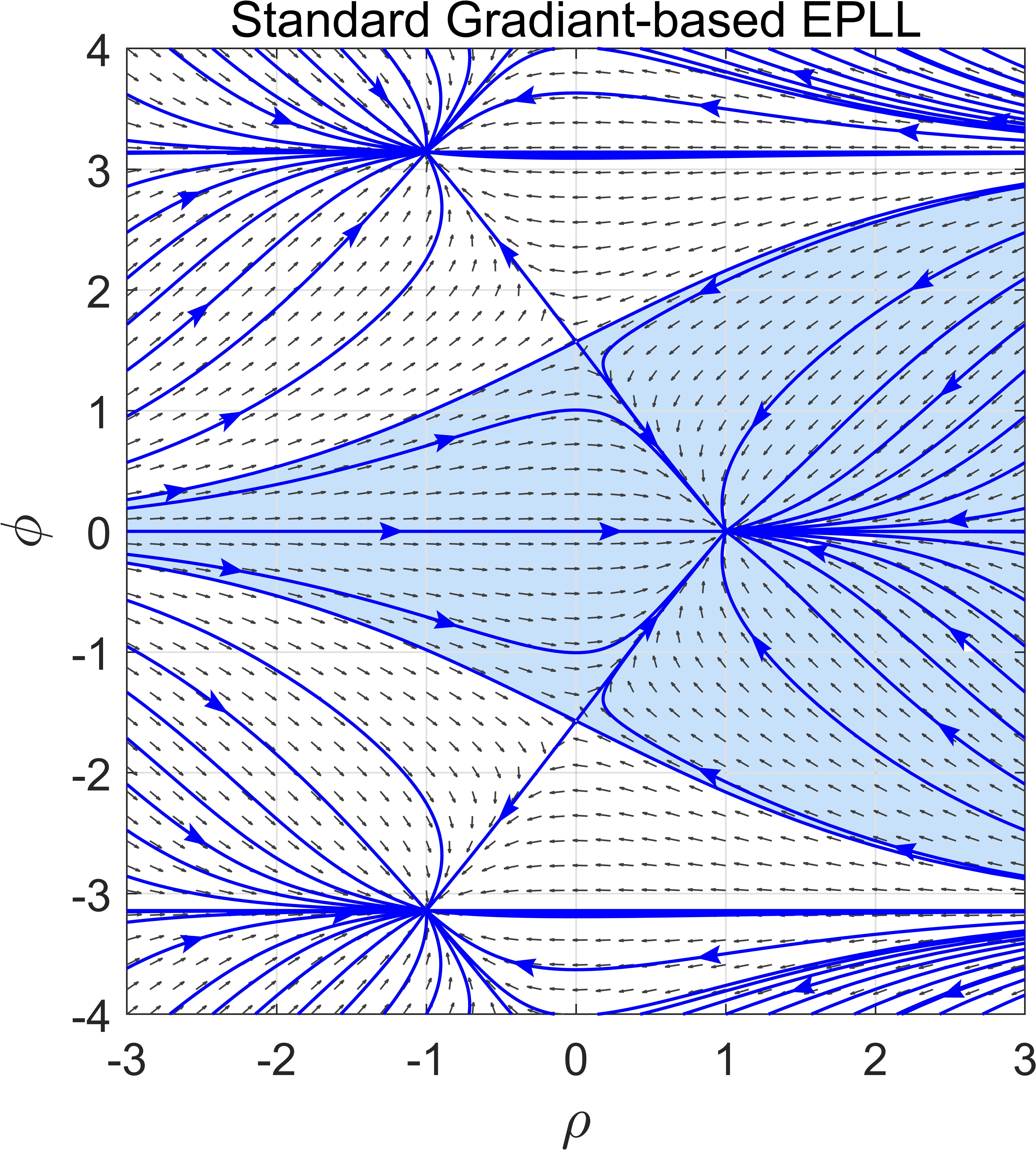}
\caption{Phase portrait of the autonomous \texttt{SEPLL}.}
\label{Fig_QuiverPlot_V2_GradientFlow}
\end{figure}

The corresponding gradient flow is
\begin{equation}
\label{EqGradientsofV2}
\dot{\boldsymbol{\theta}}
=
-\frac{\mu}{2}\nabla V
=
\mu
\begin{bmatrix}
\rho_n\cos(\phi_n-\phi)-\rho\\
\rho\rho_n\sin(\phi_n-\phi)
\end{bmatrix}
=
\begin{bmatrix}
\dot{\rho}\\
\dot{\phi}
\end{bmatrix}.
\end{equation}
Assuming \(\rho_n=1\) and \(\phi_n=2\pi\), the corresponding phase portrait is shown in Fig.~\ref{Fig_QuiverPlot_V2_GradientFlow}.

Figure~\ref{Fig_QuiverPlot_V2_GradientFlow} shows smooth convergence toward the equilibrium at
\((\rho_n,\phi_n)\), with vector magnitudes decreasing as the equilibrium is approached. The phase portrait reveals the following properties:
\begin{enumerate}
\item The trajectories converge to the desired equilibrium points
\((\rho^*,\phi^*)
=
(1,\pm2\pi n)\),
where \(n\) is an integer.

\item Degenerate equilibrium points also exist at
\((\rho^*,\phi^*)
=
(-1,\pm\pi m)\),
where \(m\) is an odd integer. More generally,
\((\rho^*,\phi^*)
=
\bigl(\rho_n(-1)^k,\,
\phi_n+k\pi\bigr)\),
where \(k\in\mathbb Z\).

\item The gradient field does not always direct trajectories toward the equilibrium along the shortest path. Instead, trajectories tend to align with characteristic incoming and outgoing straight-line paths,
whose intersections define the saddle points
\(\big(0,\,
\phi_n+\tfrac{\pi}{2}+k\pi\big),
\,
k\in\mathbb Z\).
This causes slower convergence because trajectories follow these paths before approaching the desired equilibrium.

\item These observations agree with the behavior exhibited by the original \texttt{SEPLL} in (\ref{EqKarimiStandardEPLL}).
\end{enumerate}

\subsection{Stationary Points of \texttt{NEPLL}}

The \texttt{NEPLL} was proposed to improve the relatively slow convergence of the \texttt{SEPLL} \cite{karimi2012derivation}. However, the same study reported ``undesired oscillations'' and ``convergence to spurious solutions.'' These effects were attributed to the non-quadratic nature of the cost function \(V\), whose Hessian may become indefinite, resulting in unstable Newton updates.

To better understand this behavior, we analyze the autonomous cost function \(V\) using Newton's method and examine its phase portrait. Having already derived the gradient, the corresponding Hessian is
\begin{equation}
\boldsymbol{H}
=
\begin{bmatrix}
\dfrac{\partial^2V}{\partial\rho^2}
&
\dfrac{\partial^2V}{\partial\phi\,\partial\rho}
\\[1mm]
\dfrac{\partial^2V}{\partial\rho\,\partial\phi}
&
\dfrac{\partial^2V}{\partial\phi^2}
\end{bmatrix}
=
2
\begin{bmatrix}
1 &
\rho_n\sin\Delta_\phi
\\
\rho_n\sin\Delta_\phi &
\rho_n\rho\cos\Delta_\phi
\end{bmatrix},
\end{equation}
where
\(
\Delta_\phi:=\phi-\phi_n.
\)

The resulting autonomous Newton flow is
\begin{align}
\label{EqV2HessianUpdateEPLL}
\dot{\boldsymbol{\theta}}
&=
-\mu\,
\boldsymbol{H}^{-1}
\nabla V
\nonumber\\
&=
\begin{bmatrix}
\dfrac{\mu\rho\bigl(\rho_n-\rho\cos(\phi_n-\phi)\bigr)}
{\rho\cos(\phi_n-\phi)-\rho_n\sin^2(\phi_n-\phi)}
\\[3mm]
\dfrac{\mu\rho_n\sin\!\left(2(\phi_n-\phi)\right)}
{\rho\cos(\phi_n-\phi)-\rho_n\sin^2(\phi_n-\phi)}
\end{bmatrix}
=
\begin{bmatrix}
\dot{\rho}\\
\dot{\phi}
\end{bmatrix}.
\end{align}

Assuming
\(
\rho_n=1
\)
and
\(
\phi_n=2\pi,
\)
the corresponding phase portrait is shown in Fig.~\ref{Fig_QuiverPlot_V2_NewtonFlow} (see Appendix~A for the MATLAB code used to generate the figure). Several observations can be made.

\begin{enumerate}
\item In addition to the desired minima at
\(
(1,\pm2\pi n)
\)
and the degenerate minima at
\(
(-1,\pm\pi m),
\)
where \(n\in\mathbb Z\) and \(m\) is odd, the autonomous \texttt{NEPLL} introduces spurious minima at
\(\big(0,\,
\tfrac{\pi}{2}\pm\ell\pi
\big),
\,
\ell\in\mathbb Z\),
which are indicated by the black circles in Fig.~\ref{Fig_QuiverPlot_V2_NewtonFlow}.

 \item   The unwanted second-order oscillations arise from the phase update, which is proportional to \(\sin\left(2(\phi_n - \phi)\right)\).

\item The appearance of the spurious minima is a direct consequence of the loss of positive definiteness of the Hessian. Specifically,
\[
\det(\boldsymbol{H})
=
4\rho_n
\left(
\rho\cos(\phi_n-\phi)
-
\rho_n\sin^2(\phi_n-\phi)
\right),
\]
whose zero level set coincides exactly with the denominator of the Newton flow in (\ref{EqV2HessianUpdateEPLL}). Along these singular curves, the Hessian becomes non-invertible, causing the Newton direction to become ill-defined and altering the local geometry of the optimization landscape. Consequently, saddle points of the original cost function become attractive under the Newton dynamics, giving rise to the spurious equilibrium points observed in Fig.~\ref{Fig_QuiverPlot_V2_NewtonFlow}. This behavior is consistent with the well-known limitation of Newton's method discussed by Dauphin \emph{et al.}~\cite{dauphin2014identifying}:
\begin{quote}
\small
``Second-order methods, like the Newton method, are designed to rapidly descend plateaus surrounding local minima by multiplying the gradient steps with the inverse of the Hessian matrix. However, the Newton method does not treat saddle points appropriately; \ldots\ saddle points become attractive under Newton dynamics.''
\end{quote}

\item The shaded region in the phase portrait highlights another limitation of the \texttt{NEPLL}: the basin of attraction of the desired local minimum is relatively small. Consequently, successful convergence depends critically on the initial conditions, whereas trajectories initialized outside this basin are attracted to spurious stationary points.

\item These observations closely match the behavior exhibited by the original \texttt{NEPLL} in (\ref{EqEPLLNewton}).
\end{enumerate}

\begin{figure}[htbp]
\centering
\includegraphics[width=0.73\linewidth]{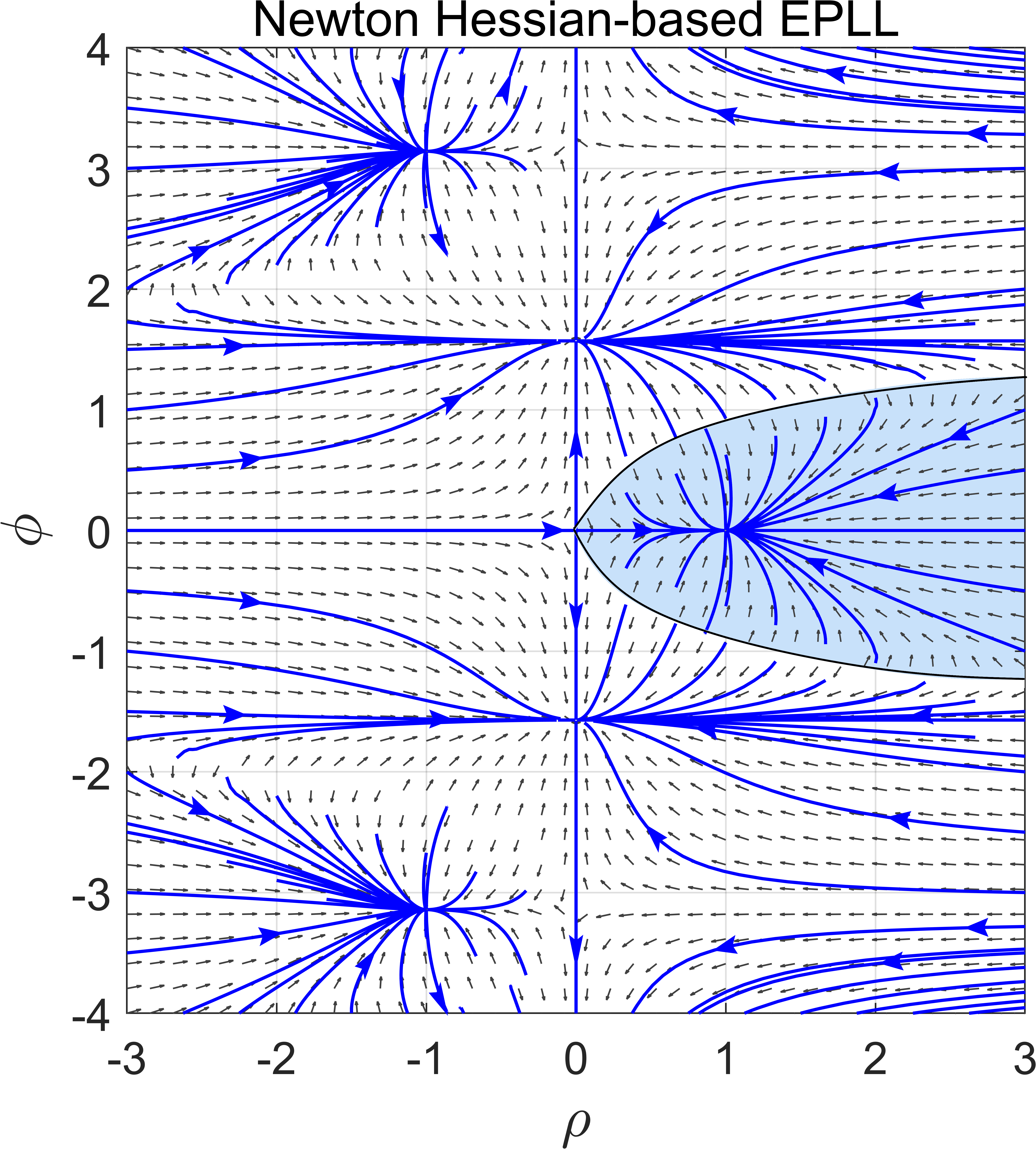}
\caption{Phase portrait of the autonomous \texttt{NEPLL}.}
\label{Fig_QuiverPlot_V2_NewtonFlow}
\end{figure}

\subsection{Stationary Points of \texttt{MEPLL}}

The autonomous \texttt{MEPLL}'s phase update differs from that of the \texttt{SEPLL}, and is given by
\begin{equation}
\label{EqAve}
\begin{aligned}
\dot{\boldsymbol{\theta}}
&=
\begin{bmatrix}
\dot{\rho}\\
\dot{\phi}
\end{bmatrix}
=
-\frac{\mu}{2}
\begin{bmatrix}
1 & 0\\
0 & \rho^2
\end{bmatrix}^{-1}
\nabla V
\\
&=
\mu
\begin{bmatrix}
\rho_n\cos(\phi_n-\phi)-\rho\\
\rho^{-1}\rho_n\sin(\phi_n-\phi)
\end{bmatrix}.
\end{aligned}
\end{equation}

\begin{figure}[t!]
\centering
\includegraphics[width=0.73\linewidth]{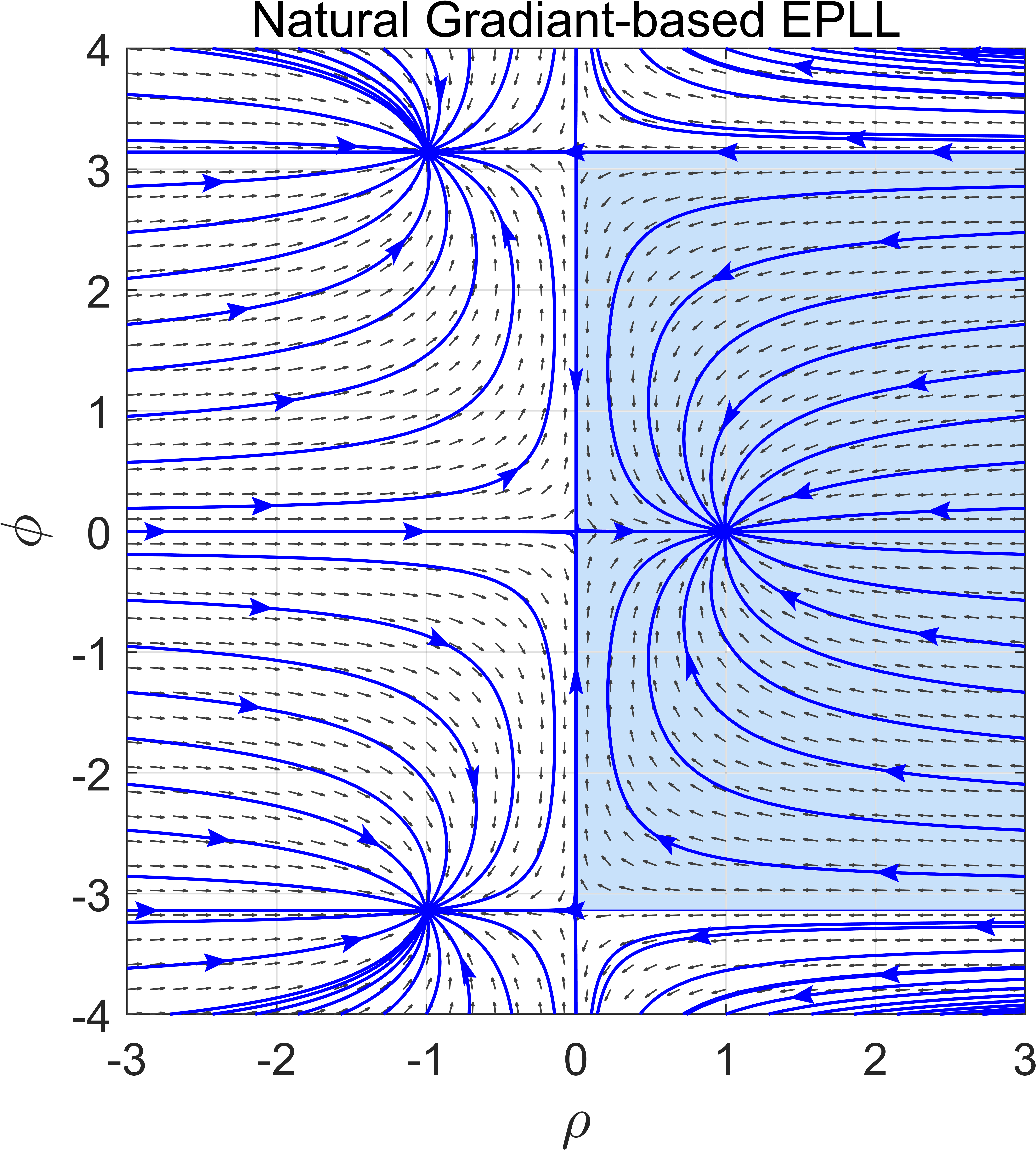}
\caption{Phase portrait of the autonomous \texttt{MEPLL}.}
\label{figQuiverplot}
\end{figure}

Assuming
\(
\rho_n=1,
\)
\(
\phi_n=2\pi,
\)
and
\(
\mu=1,
\)
the corresponding phase portrait is shown in Fig.~\ref{figQuiverplot}. The following observations can be made.

\begin{enumerate}
\item The trajectories converge smoothly to the desired equilibrium points
\((\rho^*,\phi^*)
=
(1,\pm2\pi n),
\,
n\in\mathbb Z\),
demonstrating that the autonomous \texttt{MEPLL} possesses the desired stable nodes.

\item Similar to the \texttt{SEPLL}, degenerate equilibrium points also exist at
\((\rho^*,\phi^*)
=
(-1,\pm\pi m)\),
where \(m\) is an odd integer. These equilibrium points correspond to negative amplitude estimates and can be readily detected and corrected during practical implementation, as discussed in \cite[Sec.~2.5]{KarimiBook}. Consequently, if the system is initialized with a non-positive value of \(\rho\) and no corrective action is taken, the trajectories converge to the corresponding degenerate minimum rather than the desired equilibrium. 

\item Unlike the autonomous \texttt{NEPLL}, the phase portrait contains neither spurious local minima nor saddle points. Consequently, the trajectories are not diverted toward undesired stationary points and converge directly to the desired equilibrium without the second-order oscillations observed in the Newton-based implementation.

\item The corresponding phase portrait exhibits a significantly larger basin of attraction around the desired equilibrium than those of the \texttt{SEPLL} and \texttt{NEPLL}. This enlarged basin, together with the absence of spurious stationary points, explains the faster and more reliable convergence of the \texttt{MEPLL}.
\end{enumerate}

\section{Overview of the Ahmadi--Chaudhry--Zhang Higher-Order Newton Method}

Classical Newton's method is widely used for unconstrained optimization because of its quadratic local convergence and its effective use of second-order curvature information. However, it relies only on first- and second-order derivatives and assumes that the Hessian is positive definite. Near saddle points or in nonconvex regions, the Hessian may become singular or indefinite, causing the iteration to diverge or converge to undesired stationary points.

To overcome these limitations, Ahmadi, Chaudhry, and Zhang proposed a higher-order Newton framework that augments the third-order Taylor approximation with an optimally chosen convexification term. The resulting method admits a closed-form higher-order Newton update while retaining polynomial computational complexity per iteration.

Consider a sufficiently smooth cost function
\(f:\mathbb{R}\rightarrow\mathbb{R},\)
and define the Newton correction
\(p=x-x_k.\)
The third-order Taylor approximation about the current iterate is
\begin{equation}
\label{EqThirdOrderTaylor}
T_3(p)
=f(x_k)
+f'(x_k)p
+\tfrac12f''(x_k)\,p^2
+\tfrac16f'''(x_k)\,p^3
\end{equation}

The first derivative determines the local descent direction, the second captures the local curvature, and the third describes how that curvature changes. Together, they provide a richer local approximation than the quadratic model used in classical Newton's method. The Newton correction is obtained by solving
\begin{equation}
f'(x_k)+f''(x_k)p=0,
\end{equation}
which gives the Newton update:
\begin{equation}
\label{EqNewtonStep}
p=-\frac{f'(x_k)}{f''(x_k)}.
\end{equation}
Differentiating (\ref{EqThirdOrderTaylor}) yields
\begin{subequations}
\begin{align}
T_3'(p)
&=
f'(x_k)
+
f''(x_k)p
+
\tfrac12f'''(x_k)p^2,
\\
\label{EqSecondDerivativeTaylor}
T_3''(p)
&=
f''(x_k)
+
f'''(x_k)p.
\end{align}
\end{subequations}

Unlike the quadratic model used in classical Newton's method, the Hessian of the cubic Taylor approximation depends on the optimization variable. As a result, it may become indefinite, causing the local approximation to lose convexity. To overcome this difficulty, Ahmadi \emph{et al.} augmented the cubic Taylor model with the quartic convexification term
\begin{equation}
\label{EqConvexifiedTaylor}
\Psi(p)
=T_3(p)
+\eta\,p^4,
\end{equation}
where \(\eta>0\) is the convexification parameter. The corresponding gradient and Hessian are
\begin{subequations}
\label{EqConvexifiedHessian}
\begin{align}
\Psi'(p)
&=
f'(x_k)
+f''(x_k)p
+\tfrac12f'''(x_k)p^2
+4\,\eta\,p^3
\\
\Psi''(p)
&=
f''(x_k)
+f'''(x_k)p
+12\,\eta \,p^2
\end{align}
\end{subequations}

Naturally, we would like to restore convexity without changing the original Taylor model more than necessary. To achieve this, Ahmadi \emph{et al.} selected the smallest value of \(\eta\) that guarantees a nonnegative Hessian. Since (\ref{EqConvexifiedHessian}) is quadratic in \(p\), this is accomplished by requiring its discriminant to vanish,
\begin{equation}
\label{EqDiscriminant}
\left(f'''(x_k)\right)^2
-48\,\eta\,f''(x_k)=0,
\end{equation}
yielding
\begin{equation}
\label{EqOptimalT}
\eta^\ast=\tfrac{1}{48}\frac{\left(f'''(x_k)\right)^2}
{f''(x_k)}.
\end{equation}
This gives the smallest convexification parameter that guarantees a globally nonnegative Hessian while introducing the least possible modification to the original Taylor approximation.

The stationary point satisfies
\(\Psi'(p)=0.\)
Substituting (\ref{EqOptimalT}) transforms the resulting cubic equation into a translated perfect cube, leading to the explicit solution
\(p=\beta+\sqrt[3]{\gamma}\)
(see Appendix~B for the derivation), where \(\beta\) and \(\gamma\) depend only on the first three derivatives of \(f\). Expressed directly in terms of these derivatives, the higher-order Newton correction becomes
\begin{equation}
\label{EqAhmadiUpdate}
p=-2\frac{f''(x_k)}
{f'''(x_k)}-\frac{\sqrt[3]{12f'(x_k)f''(x_k)f'''(x_k)-8\left(f''(x_k)\right)^3}}
{f'''(x_k)}.
\end{equation}
where $f'''(x_k)\neq 0$. The next iterate is therefore given by
\[
x_{k+1}=x_k+p.
\]

Equation (\ref{EqAhmadiUpdate}) defines a discrete-time optimization algorithm. Here, however, our goal is to develop a continuous-time adaptive law for the EPLL. Following the standard interpretation of gradient and Newton methods, we associate the discrete search direction with a continuous-time vector field having the same direction. This allows the resulting dynamics to be studied using phase portraits, providing insight into their stationary points, stability, and convergence behavior. Accordingly, we define the \textit{higher-order Newton flow} as
\begin{subequations}
\label{EqAhmadiFlow}\begin{align}
\dot{x}&=-2\frac{f''(x)}
{f'''(x)}-\frac{\sqrt[3]{12f'(x)f''(x)f'''(x)-8\left(f''(x)\right)^3}}
{f'''(x)}.\\
&=:-g(x)\end{align}
\end{subequations}
The ordinary differential equation (\ref{EqAhmadiFlow}) is called \textit{time homogeneous} since $g(\cdot)$ does not depend upon time $t$ \cite{meyn2022control}. 

\section{Design of a New EPLL Using \\ ACZ Higher-Order Newton Method}

To investigate the application of the Ahmadi--Chaudhry--Zhang higher-order Newton framework to the following autonomous cost, we first optimize with respect to the phase variable while treating the amplitude $\rho$ as constant. The proposed approach may be viewed as a continuous-time coordinate optimization strategy in which different optimization principles are employed along different coordinate directions according to the local geometry.
\begin{align*}
V(\rho,\phi)
:=
\rho^2+\rho_n^2
-
2\rho\rho_n
\cos(\phi-\phi_n).
\end{align*}

For notational simplicity, let
\(\Delta=\phi-\phi_n\).
The first four derivatives of $V(\rho,\phi)$ with respect to $\phi$ are
\begin{subequations}
\begin{align}
V'(\phi)
&=
2\rho\rho_n\sin\Delta,
\\
V''(\phi)
&=
2\rho\rho_n\cos\Delta,
\\
V'''(\phi)
&=
-2\rho\rho_n\sin\Delta,
\\
V''''(\phi)
&=
-2\rho\rho_n\cos\Delta.
\end{align}
\end{subequations}

Substituting these derivatives into the higher-order Newton flow gives
\begin{equation}
\label{EqAhmadiOriginal}
\dot\phi=-2\frac{V''(\phi)}
{V'''(\phi)}
-\frac{\sqrt[3]{12V'(\phi)V''(\phi)V'''(\phi)-8\left(V''(\phi)\right)^3}}{V'''(\phi)}
\end{equation}
Substituting the derivatives into (\ref{EqAhmadiOriginal}) gives
\begin{align}
\dot\phi
&=
2\cot(\phi-\phi_n)
-
\sqrt[3]
{
12\cot(\phi-\phi_n)
+
8\cot^3(\phi-\phi_n)
}.
\end{align}
The above derivation naturally raises the question of whether the same higher-order framework can also be applied to the amplitude update. Since
\[
\frac{\partial V}{\partial\rho}
=
2\bigl(\rho-\rho_n\cos(\phi-\phi_n)\bigr),
\]
and
\[
\frac{\partial^2V}{\partial\rho^2}=2,
\qquad
\frac{\partial^3V}{\partial\rho^3}=0,
\qquad
\frac{\partial^4V}{\partial\rho^4}=0,
\]
all derivatives of order three and higher vanish identically. Consequently, the higher-order correction also vanishes, reducing the third-order Taylor approximation to the quadratic model. Since $V$ is exactly quadratic with respect to $\rho$, second-order information completely characterizes its local geometry. Therefore, the proposed higher-order framework offers no additional benefit for the amplitude update and is applied only to the phase variable. The amplitude flow thus remains identical to that of the autonomous \texttt{SEPLL} and \texttt{MEPLL}. Accordingly, we propose the following autonomous higher-order Newton-based EPLL (\texttt{HOEPLL}):
\begin{subequations}
\begin{align}
\dot\rho(t)
&=
\rho_n\cos(\phi_n-\phi(t))
-
\rho(t),
\\
\dot\phi(t)
&=
2\mu
\left[
\cot(\phi(t)-\phi_n)
-
|\zeta(t)|^{1/3}
\operatorname{sign}(\zeta(t))
\right],
\\
\nonumber
\text{where}~~
\zeta(t)
&=
\tfrac{3}{2}
\cot(\phi(t)-\phi_n)
+
\cot^3(\phi(t)-\phi_n).
\end{align}
\end{subequations}

\begin{figure}[b!]
\centering
\includegraphics[width=0.73\linewidth]{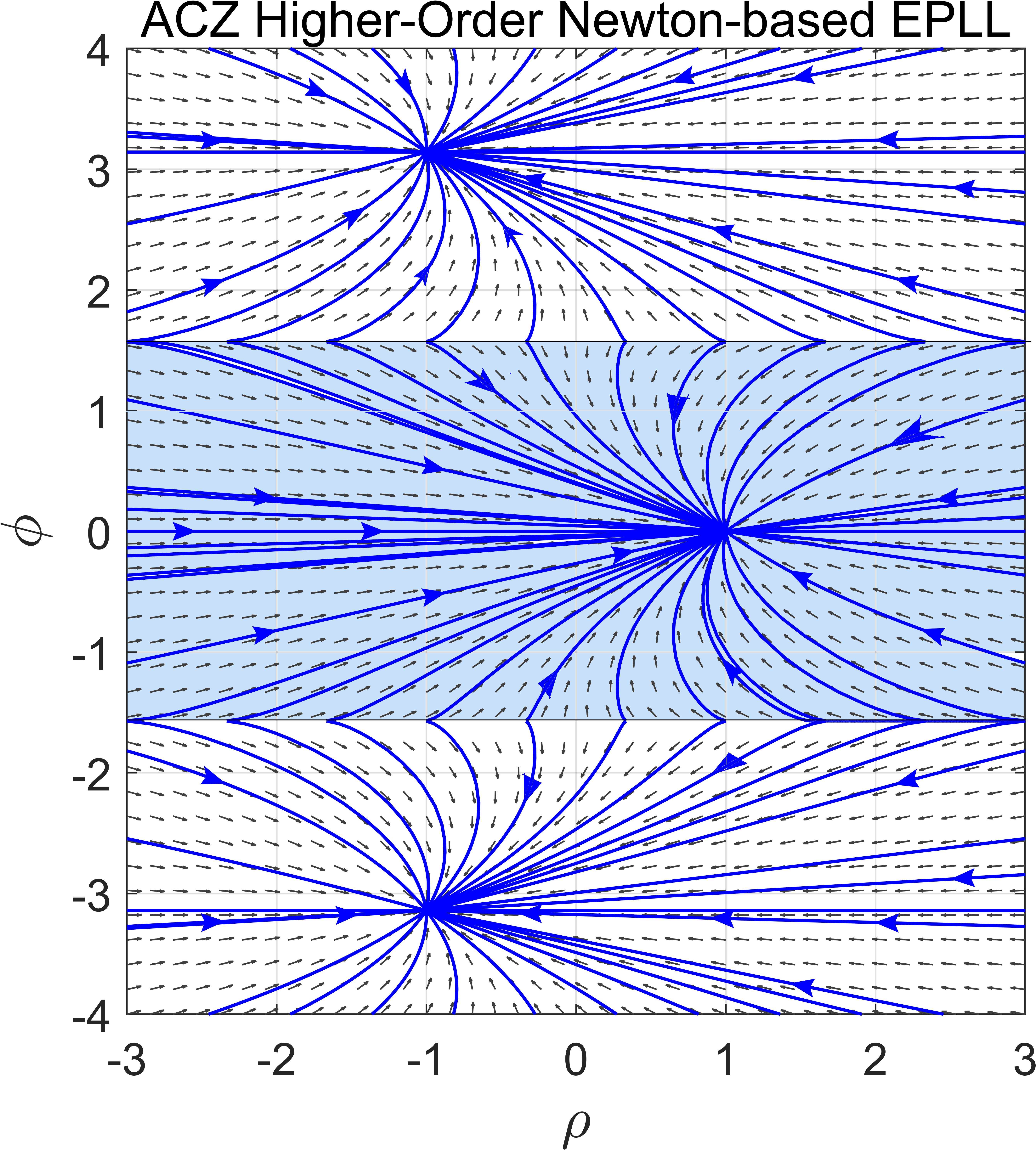}
\caption{Phase portrait of the autonomous \texttt{HOEPLL}.}
\label{Fig_QuiverPlot_HOEPLL}
\end{figure}

Assuming \(\rho_n=1\), \(\phi_n=2\pi\), and \(\mu=1\), the corresponding phase portrait of the proposed autonomous higher-order Newton-based EPLL (\texttt{HOEPLL}) is shown in Fig.~\ref{Fig_QuiverPlot_HOEPLL}. Several important observations can be made.

\begin{enumerate}

\item The phase portrait exhibits the expected periodicity with respect to the phase variable. In particular, the flow repeats over successive intervals of length \(2\pi\), with the desired equilibrium points located at
\[
(\rho^*,\phi^*)
=
(\rho_n,\phi_n+2k\pi),
\qquad
k\in\mathbb Z.
\]
Consequently, it is sufficient to analyze the principal interval
\[
-\tfrac{\pi}{2}
<
\phi-\phi_n
<
\tfrac{\pi}{2},
\]
since the remaining dynamics follow by periodicity.

\item Within the principal phase interval, all trajectories converge to the unique desired equilibrium
\((\rho^*,\phi^*)
=
(\rho_n,\phi_n)\),
indicating a remarkably large basin of attraction. In particular, convergence is achieved even when the initial amplitude estimate is negative, provided that the initial phase satisfies
\(-\tfrac{\pi}{2}
<
\phi-\phi_n
<
\tfrac{\pi}{2}\).
Unlike the autonomous \texttt{MEPLL}, the proposed flow therefore naturally steers the trajectories toward the physically meaningful positive-amplitude equilibrium over a broad range of initial conditions.

\item No spurious local minima are observed anywhere in the plotted region. Consequently, trajectories are not diverted toward undesired stationary points, eliminating one of the principal shortcomings of the autonomous \texttt{NEPLL}.

\item Likewise, no saddle points or saddle-attraction phenomena are observed in the plotted region. The vector field remains smooth throughout the state space, and the trajectories converge directly toward the desired equilibrium without the separatrices and competing attraction regions observed in the autonomous \texttt{SEPLL} and \texttt{NEPLL}.

\item The trajectories approach the desired equilibrium smoothly from a wide range of initial conditions. The corresponding vector field appears well directed throughout the plotted region, suggesting favorable convergence characteristics. 

\item Compared with the autonomous \texttt{SEPLL}, \texttt{NEPLL}, and \texttt{MEPLL}, the proposed \texttt{HOEPLL} possesses the largest observed basin of attraction around the desired equilibrium. The combination of a large attraction region, the absence of spurious stationary points, and the apparent elimination of saddle-attraction phenomena suggests significantly improved robustness with respect to the initial conditions, making the proposed higher-order Newton flow particularly attractive for practical EPLL implementations.

\end{enumerate}

Since the proposed higher-order EPLL differs from the modified EPLL only in the phase adaptation law and exhibits similar estimation behavior under representative power-system conditions, a comprehensive simulation study in practical power-system scenarios is omitted in this paper. Such an investigation, including robustness under harmonics, frequency deviations, voltage disturbances, and grid faults, will be presented in future work.

\section{Conclusions}

This paper presented a new higher-order enhanced phase-locked loop (\texttt{HOEPLL}) based on the recently proposed Ahmadi--Chaudhry--Zhang (ACZ) higher-order Newton framework. A continuous-time interpretation of the ACZ method was developed, establishing a connection between higher-order optimization and adaptive synchronization systems. A coordinate optimization strategy was adopted in which the higher-order Newton flow was applied only to the phase update, while the amplitude update retained its classical form because the corresponding cost function is exactly quadratic.

The proposed algorithm was analyzed through autonomous phase portraits and compared with the standard, Newton, and modified EPLL formulations. The analysis showed that the proposed flow eliminates the spurious equilibria and saddle-attraction phenomena of the autonomous Newton EPLL while exhibiting a substantially larger basin of attraction. These results demonstrate that higher-order optimization provides a systematic framework for designing adaptive phase-locked loops with improved convergence properties and may also prove useful for other nonlinear adaptive estimation and synchronization algorithms.

\appendices

\section{MATLAB Code for the Phase Portrait of \texttt{NEPLL}} 

The \texttt{NEPLL} update contains singularities, which require careful numerical treatment during phase-portrait generation. To facilitate reproducibility and assist readers, the MATLAB code used to generate the phase portrait is included below:

\begin{lstlisting}[style=Matlab-editor,basicstyle=\color{black}\ttfamily\footnotesize]
rhon = 1;  phin = 2*pi;
Dmin = 5e-2; % floor on denominator
odefun = @(t,x) local_field(x, rhon, phin, Dmin);
progendtime = 25;
plotpp(odefun,'tspan', progendtime,...
    'plotNonSaddleTrajectory',true, ...
    'plotEPs',false,'plotQuiver',true,...
    'quivercolor', [1,1,1]*0.25,...
    'linecolor', [0,0,1],...
    'arrowDensity',3, 'xPlotNum',15,...
    'yPlotNum',15, 'arrowsize',10,...
    'quiverDensity',40, 'xlim', [-3, 3], 'ylim', [-4, 4]); hold on; 
ylabel('$$\phi$$','interpreter', 'latex');
xlabel('$$\rho$$','interpreter', 'latex'); 
% Adding trajectories inside (optional)
rho0 = linspace(-3,3,10); % Initial rho values
for k = 1:length(rho0)
    % phi = pi/2
    [~,x] = ode45(odefun,[0 progendtime],...
        [rho0(k); pi/2+0.1]);
    plot(x(:,1),x(:,2),'b','LineWidth',1)
    [~,x] = ode45(odefun,[0 progendtime],...
        [rho0(k); pi/2-0.1]);
    plot(x(:,1),x(:,2),'b','LineWidth',1)
    % phi = -pi/2
    [~,x] = ode45(odefun,[0 progendtime],...
        [rho0(k); -pi/2+0.1]);
    plot(x(:,1),x(:,2),'b','LineWidth',1)
    [~,x] = ode45(odefun,[0 progendtime],...
        [rho0(k); -pi/2-0.1]);
    plot(x(:,1),x(:,2),'b','LineWidth',1)
end
% Handling potential singularities.
function dx = local_field(x, rhon, phin, Dmin)
    d = x(2) - phin;
    D = x(1)*cos(d) - rhon*sin(d)^2;
    s = sign(D);
    if s == 0, s = 1; end
    % clamp |D| away from 0, KEEP its sign
    Dreg = s * max(abs(D), Dmin);     
    dx = [ x(1)*(rhon - x(1)*cos(d));
          -rhon*sin(2*d) ] / Dreg;
end
\end{lstlisting}

\section{Translated Perfect-Cube Representation}

Substituting the optimal convexification parameter (\ref{EqOptimalT})
into (\ref{EqConvexifiedHessian}a), the stationarity condition
\(
\Psi'(p)=0
\)
becomes
\begin{align*}
0
=
f'(x_k)
+
f''(x_k)p
+
\tfrac12f'''(x_k)p^2
+
\tfrac{1}{12}\frac{\left(f'''(x_k)\right)^2}
{f''(x_k)}
p^3.
\end{align*}
Factoring out the coefficient of the cubic term gives
\begin{align*}
0=
\frac{\left(f'''(x_k)\right)^2}
{12f''(x_k)}\!
\Bigg[
p^3
+
\frac{6f''(x_k)}
     {f'''(x_k)}
p^2
+
\frac{12\left(f''(x_k)\right)^2}
     {\left(f'''(x_k)\right)^2}
p
\Bigg]\!
+\!
f'(x_k)
\end{align*}
Now observe that
\begin{align*}
\left(
p+
2\frac{f''(x_k)}
        {f'''(x_k)}
\right)^3
\!=
p^3
& +
6
\frac{f''(x_k)}
     {f'''(x_k)}
p^2
+ \\ \nonumber &\,\,
12
\frac{\left(f''(x_k)\right)^2}
     {\left(f'''(x_k)\right)^2}
p+ 8 \frac{\left(f''(x_k)\right)^3}
     {\left(f'''(x_k)\right)^3}.
\end{align*}
Hence,
\begin{align*}
\nonumber p^3
+
6
\frac{f''(x_k)}
     {f'''(x_k)}
p^2
+
& 12
\frac{\left(f''(x_k)\right)^2}
     {\left(f'''(x_k)\right)^2}
p\\
& =
\left(
p+
2
\frac{f''(x_k)}
     {f'''(x_k)}
\right)^3
-
8
\frac{\left(f''(x_k)\right)^3}
     {\left(f'''(x_k)\right)^3}.
\end{align*}
Substituting this identity into the stationarity equation yields
\begin{align*}
0=
\frac{\left(f'''(x_k)\right)^2}
{12f''(x_k)}
\left(
p+
2
\frac{f''(x_k)}
     {f'''(x_k)}
\right)^3
+
f'(x_k)
-
\tfrac23
\frac{\left(f''(x_k)\right)^2}
     {f'''(x_k)}.
\end{align*}
From which, the closed-form higher-order Newton correction (\ref{EqAhmadiUpdate}) follows directly.

\bibliographystyle{unsrt}
\bibliography{references}

\end{document}